\documentclass[prl,twocolumn,amsmath,amssymb,superscriptaddress]{revtex4}

\usepackage{graphicx}
\usepackage{dcolumn}
\usepackage{amsmath}
\usepackage{amssymb}

\newcommand{\be}{\begin{equation}}
\newcommand{\ee}{\end{equation}}
\newcommand{\bea}{\begin{eqnarray}}
\newcommand{\eea}{\end{eqnarray}}
\newcommand{\p}{\partial}

\begin{document}

\title{Stochastic Loewner evolution for
conformal field theories with Lie group symmetries}

\author{E. Bettelheim}
\affiliation{James Frank Institute, University of Chicago, 5640 S.
Ellis Ave. Chicago IL 60637}
\author{I. A. Gruzberg}
\affiliation{James Frank Institute, University of Chicago, 5640 S.
Ellis Ave. Chicago IL 60637}
\author{A. W. W. Ludwig}
\affiliation{Department of Physics, University of California,
Santa Barbara, CA 93106}
\author{P. Wiegmann}
\affiliation{James Frank Institute, University of Chicago, 5640 S.
Ellis Ave. Chicago IL 60637}

\begin{abstract}
The stochastic Loewner evolution is a recent tool in the study of
two-dimensional critical systems. We extend this approach to the
case of critical systems with continuous symmetries, such as SU(2)
Wess-Zumino-Witten models, where domain walls carry an additional
spin 1/2 degree of freedom.
\end{abstract}

\maketitle

\paragraph*{ Introduction.}


Traditionally, critical phenomena are described by scale invariant
fluctuations of local order parameters.
In two dimensions,
statistical
mechanics models
and conformal field theories
(CFT) \cite{BPZ}, describing their critical behavior, can often be
formulated in terms of fluctuating loops --- simple critical
curves. These curves can be viewed as external perimeters of
critical clusters\cite{nienhuis}.

A radically new development, called stochastic Loewner evolution
(SLE) \cite{Schramm:SLE,Lawler book,Cardy Review}, revitalizes the
latter representation of critical
models
in two dimensions,
addressing directly the stochastic geometry of critical curves.
SLE suggests a specific  description of
the statistics of
the
critical curves through
simple
Brownian motion.

So far the applications of the SLE approach were limited to the
least structured CFTs with
central charge $c \leqslant 1$
\cite{Bauer:2002tf}. Yet, many applications of CFT, including
condensed matter problems, possess continuous internal symmetries,
such as e.g.  SU(2) spin-rotational symmetry for electrons in a
solid. The most important CFTs with such symmetries are
Wess-Zumino-Witten (WZW) models \cite{Knizhnik:Zamolodchikov},
whose central charge is  $c \geqslant 1$.
Popular applications
include
spin chains~\cite{SpinChains}, Kondo problems of a magnetic
impurity in metal \cite{Nozieres+Blandin}, and D-branes
\cite{Alekseev:1998mc}.

Can the SLE approach describe more structured CFTs such as WZW
models? Here we address this question.
Indeed,
some
 WZW models {\it
at
level one\/} can
also be represented by fluctuating
loops, but now the loops are decorated by a representation of the
Lie algebra  \cite{kondev}.

In this paper we show that the SU(2) WZW model, {\it at any
level}, can be described by a composition of the standard SLE
stochastic
process and a Brownian motion in the  Lie algebra
\cite{Rasmussen}. Being interesting by itself, this
representation
allows one
 in particular
to compute, amongst other properties, the
fractal geometry
of the loops (we report this result elsewhere).



\paragraph*{Stochastic Loewner evolution.}

Consider a critical (scale invariant) system in the upper half
complex plane, called
the
{\it physical plane}. We impose
different boundary conditions to the left and to the right of  a
point $w_0$
on
 the real axis,
chosen so that a domain wall emanates
from $w_0$. The domain wall is a fluctuating curve. SLE
interprets this curve as the trace of a self-avoiding walk
progressing with a properly chosen time $t$.
\begin{figure}[t]
\includegraphics[width=2in]{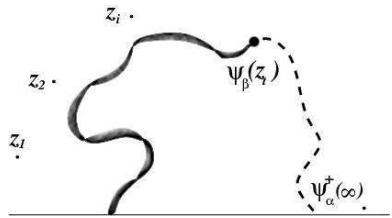}
\caption{ A ribbon represents an SLE trace carrying spin degrees
of freedom. Points $z_i$ are positions of primary fields. A
spin-1/2 operator positioned at the tip extends the trace up to
another spin-1/2 operator on the boundary.}
\end{figure}
At $t = \infty$ the trace hits the boundary and surrounds a
critical domain. At any $t < \infty$ one considers the slit domain
{\bf H}$_t$, i. e. the upper half plane from which the trace is
removed (see the figure). The slit domain can be mapped
conformally onto the upper half plane by a function $f(z)$
normalized so that $f(z) = z + 2t/z + \dots$ near $z = \infty$.
The coefficient $t$ is called the capacity of the trace and is
chosen to be the time of the evolution. Under this map the tip
$z_t$ of the trace in the physical plane maps to a point $w_0 +
\xi(t)$ on the real axis. Loewner's equation connects the
evolution of the conformal map $f(z)$ to that of the image $w_0 +
\xi(t)$: $\dot f(z) = 2/(f(z) - w_0 - \xi(t))$. It is convenient
to shift the map to become $w(z) = f(z) - \xi(t)$, so that the tip
is always mapped to the fixed point $w_0$ on the real axis of the
{\it mathematical plane} (coordinate $w$). Then Loewner's equation
becomes
\begin{align}
d w(z) = \frac{2dt}{w(z) - w_0} - d\xi. \label{Loewner}
\end{align}
In SLE,
$\xi(t)$ is
 a Brownian motion: $\prec \! \dot\xi(t)
\dot\xi(0) \!\! \succ \, = \kappa\delta(t)$. We use the symbol
$\prec \!\! ... \!\! \succ$ for the {\it stochastic} average over
the Brownian motion $\xi$ not to be confused with the CFT average
$\langle ... \rangle$ as e.g.  in Eq. (\ref{spect}). Eq.
(\ref{Loewner}) generates a stochastic self-avoiding trace whose
statistics is that of a domain wall in a CFT with central charge
$c \leqslant 1$, determined by the noise strength $\kappa$ through
the relation $c = 1 - 6(\sqrt{\kappa/4} - \sqrt{4/\kappa})^2$.

\paragraph*{The WZW model} involves a field $G(w,\bar w)$
taking values in a Lie group. It is a CFT whose action is
invariant under independent holomorphic left and anti-holomorphic
right multiplication $G\to UGV^{-1}$.
It possesses
corresponding
conserved
Noether currents
$J_L=\partial G G^{-1}$,
$J_R= -G^{-1} \bar\partial G$
which
are
holomorphic ($J_L$) and
anti-holomorphic ($J_R$). This requires the matrix $G(w,\bar w)$
to be a product of two holomorphic matrices
$ G(w,\bar w) = g^{\vphantom{-1}}_L(w)g^{-1}_R(\bar w)$.
In terms of these,
the currents are expressed as $J_L = (\p
g^{\vphantom{-1}}_L) g^{-1}_L$, $J_R = (\bar\p
g^{\vphantom{-1}}_R) g_R^{-1}$.

Conformal and gauge invariant boundary conditions
require that the current normal to the boundary vanishes,
i.e. $J_L = J_R$ on the real axis \cite{Cardy1984, Ishibashi:1988kg}.
This
condition `glues'
 holomorphic and anti-holomorphic
fields,
$g_L(w)=g_R(w) \Lambda$
(for ${\rm Im} \  w=0$),
where $\Lambda$ is a matrix in a
Cartan subgroup. As a result, the field $G=g_R(w)\Lambda
g^{-1}_R(\bar w)$
belongs on the boundary
to a conjugacy class
(which, in the quantum theory, is quantized)
\cite{Alekseev:1998mc}.
For
SU(2),
conjugacy classes are
2-spheres $S^2$
parametrized
by a unit vector $\vec n$, or points.
A boundary condition can be thought of
as
being associated
with a spin\cite{Ishibashi:1988kg, AffleckLudwig1991NuclPhysB}.
A change of boundary condition at
some point on the real axis can be described
by a so-called boundary condition changing
operator\cite{CardyNPB1989}
which, in the present case, carries spin.

After these comments, consider a critical cluster of a WZW theory.
Its boundary is characterized by
its fluctuating geometry
 (the shape of the cluster), which is `decorated' by a spin. Together this can
be seen as a self-avoiding walk in the physical plane with a
fluctuating spin-1/2 degree of freedom.

The next paragraph recounts arguments which are well described in
the SLE literature\cite{Schramm:SLE,Lawler book,Bauer:2002tf}.
Therefore, we mention only briefly the main steps.

\paragraph* {Martingales and Correlation functions in the slit domain\/.}

Let us study correlation functions (conformal blocks) of primary
fields\cite{BPZ,Knizhnik:Zamolodchikov}
of a conformally invariant model, called
``spectators'', inserted at points $z_i$ in the slit domain {\bf
H}$_t$ made by the trace (see the figure). The positions $z_i$ do
not move while the trace evolves. Each field
$\phi_{\alpha_i}(z_i)$ carries  spin $s_i$ and conformal weight
$h_i$ (for a theory with no Lie group symmetry set $s_i=0$). We
denote by $\psi_\alpha$
a
boundary condition changing operator\cite{CardyNPB1989}
which is also a primary field
[for the
\rm{SU(2)} model
we choose it to be a
spin 1/2 primary field,
while for $c<1$
a $(2,1)$ or a $(1,2)$ field in
the Kac classification].
 Such operators are inserted
at the tip
and at the end (i.e. at infinity)
of the trace.

A bulk operator is the
product\cite{Cardy1984}
of two holomorphic operators
located at `Schwarz-symmetric' points as $\phi_{\alpha_i}(z_i)
\phi_{\beta_i}(z_i^*)$, and transforming in the same
representation. An example of a spectator is the matrix $G$
itself. We denote a product of spectators by ${\cal O}(\{ z_i\})$
and
their correlation function
in ${\bf H}_t$
 by
${\cal F} = \langle \alpha|
{\cal O}| \beta \rangle_{{\bf H}_t}$.  In terms of CFT
this
 correlation function
reads
\begin{align}
{\cal F}(t, \{z_i\}) = \frac{\langle \psi^\dagger_\alpha(\infty)
\, {\cal O} \, \psi_\beta (z_t )\rangle_{{\bf H}_t}}{(1/2) \langle
\psi^\dagger_\gamma(\infty) \psi_\gamma( z_t)\rangle_{{\bf H}_t}}.
\label{spect}
\end{align}

It is known\cite{Bauer:2002tf}
that
if we average
the correlator (\ref{spect}) in the slit domain
over all configurations of the SLE trace, we obtain a CFT
correlator in
the upper half-plane ${\bf H}$
with the
boundary operators inserted:
\begin{align}
\prec \! \langle \alpha|  {\cal O}|\beta\rangle_{{\bf H}_t} \!\!
\succ &= \left\langle \psi^\dagger_\alpha(\infty) \, {\cal O}\,
\psi_\beta (w_0) \right\rangle_{\bf H}. \label{stochastic-average}
\end{align}
This implies, as we now review, the steady state condition
\begin{align}
\partial_t \prec \! {\cal F} \! \succ &= 0. \label{martingale}
\end{align}

A stochastic quantity, whose average is time-independent, is known
as a `martingale'. The argument showing that a correlation
function with two boundary operators is a martingale is as follows
\cite{Bauer:2002tf}. At time $t$ we decompose the trace into two
parts: one between points $w_0$ and $z_t$, and the other between
$z_t$ and infinity (see the figure). We average over all
configurations of the trace in two steps. First, we fix the first
part and average over the second. Then we average over the first
part. The first average can  be seen as the CFT average in the
slit domain formed by the trace, with two boundary operators
inserted, one at the tip of the trace and one at infinity.  After
performing this (first) average we obtain the quantity in
(\ref{spect}). The insertion of the boundary operators effectively
averages over the second piece of the domain wall. The second
average over the shape of the first part of the trace, as on the
l.h.s. of (\ref{stochastic-average}) gives us back the original
correlator (r.h.s. of that equation).  The latter however does not
depend on the  choice of the midpoint $z_t$. Thus the stochastic
mean of the correlator ${\cal F}$ is time independent. It is a
`martingale'.

%
%

\paragraph*{Stochastic evolution on {\rm{ SU(2)}} manifold.}

In a WZW model we expect not only the geometrical fluctuations due
to the growing trace, but also
stochastic
\rm{SU(2)}
rotations.  Accordingly
we introduce
additional,
 independent Brownian motions in the left and right
su(2) Lie algebras, $\theta_{L,R}=\theta_{L,R}^a\,S^a$, with
variance
\begin{align}
{\prec \! \dot\theta_{L,R}^a(t) \, \dot\theta_{L,R}^b(0)\!\!\succ}
= \tau \delta^{ab}\delta(t).
\end{align}
Again, we use the symbol $\prec \!\! ... \!\! \succ$ for the {\it
stochastic} average over the Brownian motions $\xi$ and
$\theta^a$.
$S^a$ are generators of su(2) in a
representation conventionally normalized as $S^a S^a = s(s+1)$. We
define a stochastic evolution in the (complexified) Lie algebra by
the equations
\begin{align}
\Omega_L &= \frac{d \theta_{L }}{w - w_0}, & \Omega_R &= \frac{d
\theta_{R }}{{\bar w} - w_0}, \label{Omega}
\end{align}
where $\Omega_{L,R}=( d g^{\vphantom{-1}}_{L,R})g^{-1}_{L,R} $, $d
\theta_{L,R} = \dot\theta_{L,R} \ dt$. Here the time $t$ is the
capacity of the trace, and the time derivative is taken at a fixed
point $w$ in the mathematical plane. Under this evolution, we let
the matrix $G$ itself evolve as
$d G = \Omega_L G - G \, \Omega_R$. 
These equations respect the form of $G$ as a product of left and
right moving factors. The boundary conditions require the left and
right Brownian motions to be equal, $\theta_R = \theta_L$. From
now on we will follow only holomorphic components, dropping the
index $L$, as if there were no boundary  \cite{Cardy1984,
AffleckLudwig1991NuclPhysB}.

The pole in the evolution equation (\ref{Omega}) located at the
image of the tip of the
trace indicates
 the presence of a source
of current $J$ at $w = w_0$ in the mathematical plane. This source
originates from a juxtaposition of two different gauge invariant
boundary conditions to the left and to the right of $w_0$. We
select the pair of boundary conditions so that the domain wall,
located  in the physical plane, carries spin 1/2. In the language
of boundary CFT, this corresponds to a boundary
changing operator, transforming in the spin 1/2 representation, to
appear at position $w_0$ \cite{CardyNPB1989}. The spin 1/2 at the
tip of the trace in the physical plane fluctuates during the
evolution, and leads to a ``twisting'' of the domain wall (see the
figure).

Making use of
the
current $J(w)=\partial_w G G^{-1}$
in the
mathematical plane,
 the evolution equation (\ref{Omega}) can be rewritten in
a form where the time derivative is taken at a fixed point in the
physical plane:
\begin{align}
\Omega &= \frac{d\theta}{w(z) - w_0} + dw(z) J(w(z)).
\label{Omega-z}
\end{align}

Under an infinitesimal gauge transformation the current changes as
$dJ = [\Omega, J] + \p_w \Omega$. With the help Eqs.
(\ref{Loewner}, \ref{Omega-z}) we obtain
\begin{align}
dJ = - \frac{d\theta}{(w - w_0)^2} + \Big(\Omega_A + h
\frac{dw'}{w'} + \frac{d w}{w'} \p_z \Big) J. \label{dJ}
\end{align}
Here $\Omega_A J = [d\theta, J]/(w - w_0)$ is the adjoint action
of the evolution (\ref{Omega}), $h = 1$ is the conformal weight of
the current, and $w' = \partial_z w(z)$.
(\ref{dJ}) is a  Langevin equation for the current,
ressembling [using (\ref{Loewner})]
the
operator product expansion (OPE)
of the latter with a primary boundary operator
located at the tip.

\paragraph*{Langevin  equation.}

While all the spectator points $z_i$ in the physical plane remain
fixed under the time evolution, the trace evolves, and together
with the infinitesimal rotations this leads to a
Langevin dynamics
for
 correlators
 $\mathcal{F}$ of primary fields,
as defined in
(\ref{spect}).
Using Loewner's equation
(\ref{Loewner})
 and Eq. (\ref{Omega}),
this is conveniently written through negative grades and global
parts of Virasoro and Kac-Moody algebra generators:
\begin{align}
d{\cal F} &= \left(-d\theta^a {\cal J}^a_{-1} + d\xi \,
{\cal L}_{-1} - 2dt{\cal L}_{-2}\right){\cal F}, \label{Langevin-F} \\
{\cal L}_{-n}  &= \sum_i \left(\frac{h_i (n - 1)}{(z_i - w_0)^n} -
\frac{1}{(z_i - w_0)^{n - 1}} \frac{\partial}{\partial
z_i}\right), \quad n \geqslant -1, \nonumber \\
{\cal J}^a_{-n} &= -\sum_i \frac{S^a_i}{(z_i - w_0)^n},\quad n =
0, 1, \dots \label{KM-modes}
\end{align}
(The sum extends over all spectators and the boundary operator at
infinity.)
Similarly,
the
 Langevin equation (\ref{dJ}) for the current reads,
when written in this manner,
\begin{align}
dJ = d\theta^a {\cal J}_{-2}^a + \big(- d\theta^a{\cal J}^a_{-1} +
d\xi {\cal L}_{-1} - 2dt{\cal L}_{-2} \big) J, \label{Langevin-J}
\end{align}
where now all the generators have the form of Eq. (\ref{KM-modes})
but there is only a single term with $z_i \to z$ in the sums.

\paragraph*{Diffusion equation.}

Let
us average the correlator ${\cal F}(t)$
in (\ref{spect})
over all
configurations of the fluctuating geometry of the trace and its
spin 1/2 degree of freedom. Since the evolution has the form of a
Langevin dynamics, the expectation value obeys a diffusion
equation. The latter is obtained in the standard manner
(see for example, \cite{Bauer:2002tf}).
We average
Eq. (\ref{Langevin-F}) over the Gaussain noises.
The terms linear in $dt$ come from the first and the second order:
$\prec \! d {\cal F} \!\! \succ \, = \bigl(\prec \! {\cal G}^{-1}
d{\cal G} \!\! \succ + \frac{1}{2} \prec \! ( {\cal G}^{-1} d{\cal
G} )^2 \!\! \succ \bigr)$ $\prec \! {\cal F} \!\! \succ$. Here
$\cal{G}$ is the time evolution operator, ${\cal G}(t)$, defined by
${\cal F}(t) = {\cal G}(t)\,{\cal F}(0)$. One obtains the
diffusion equation:
\begin{align}
\partial_t \! \prec \! {\cal F} \! \succ &= -H \prec \! {\cal F} \!
\succ, \qquad {\rm where}
\label{diffusion-F}\\
H &= - \frac{\kappa}{2}{\cal L}_{-1}^{2} +2{\cal L}_{-2} -
\frac{\tau}{2} {\cal J}^a_{-1} {\cal J}^a_{-1}.
\label{Hdiffusion}
\end{align}

Similarly, we may consider correlators with insertions of the
current operator as an additional spectator. Denoting by ${\cal
F}_J = \langle \alpha| {\cal O}(\{z_i\})J(z)|\beta \rangle_{{\bf
H_t}}$ a correlator of primary fields with an insertion of the
current we obtain, with the help of (\ref{Langevin-J}), a
diffusion-type equation
\begin{align}
\p_t \prec \! {\cal F}_J \! \succ &= -H \prec \! {\cal F}_J \!
\succ - \tau {\cal J}^a_{-1} {\cal J}^a_{-2} \prec \! {\cal F} \!
\succ, \label{diffusion-J}
\end{align}
where the last (anomaly) term comes from the first term in Eqs.
(\ref{dJ}, \ref{Langevin-J}). Here ${\cal J}_{-2}^a$ acts only on
the position $z$ of the current insertion and ${\cal J}_{-1}^a$ on
all the spectators apart form the current, while the operators in
$H$ act on all spectators, including the current insertion.

\paragraph* {Singularity  at the tip.}

So far
the variances $\kappa$ and $\tau$ of the two types of
Brownian motion where treated as independent parameters. A simple
physical requirement connects them. We may be interested in
stochastic processes where  martingales
(correlation functions)
 do not have essential singularities
as a spectator approaches the tip of the trace.
In other words,  the singularities of the solutions of
the differential equations
feature only branch cuts, i.e. the equations
are Fuchsian.
This occurs only if
\begin{align}
\kappa + \tau &= 4.
\label{kappa+tau}
\end{align}
The simplest way to  obtain this condition
is to
demand that
the stochastic average of the one-point function of the current
exhibits only a single pole as the current insertion approaches
the tip of the trace. Setting ${\cal O}=1$ in (\ref{diffusion-J})
we see that the stochastic average of the current one-point
function is a zero mode of the operator
\begin{align}
\frac{\kappa}{2}\p_z^2 + \frac{\tau}{2}\frac{2}{(z - w_0)^2} +
2\p_z\frac{1}{z - w_0}.
\end{align}
The requirement that this zero mode be a single pole yields the
important condition (\ref{kappa+tau})
relating the two variances.
It implies in particular that
$0 \leqslant \kappa \leqslant 4$ since the variance $\tau$
is
non-negative; therefore,
the trace
 does not intersect itself
\cite{Lawler book}.
[If (\ref{kappa+tau}) is not satisfied, one still
appears to obtain a
possible stochastic process, but
with essential singularities in the martingales.]

\paragraph*{Knizhnik-Zamolodchikov equation and conformal weight.}

The second order differential
equation
(\ref{martingale},\ref{diffusion-F},\ref{Hdiffusion})
can
be reduced to the first order
Knizhnik-Zamolodchikov equation\cite{Knizhnik:Zamolodchikov},
as we now demonstrate.

Let us denote by $L_n$ and $J_n$ the Virasoro and the current
algebra operators acting on the
 boundary
condition changing
operator
$\psi(w_0):=$ $ \psi_\beta(w_0)$
 at the tip of the trace.
In particular,
 $L_0 = h_0$, where $h_0$ is the conformal weight of
$\psi$, $L_{-1} = \partial_{w_0}$, and $J_0^a = \sigma^a/2$ acts
on the spin 1/2 of $\psi$. The operators in (\ref{KM-modes}) are
representations of these operators. Note that $J_0^a\prec \! {\cal
F} \! \succ = {\cal J}_0^a\prec \! {\cal F} \! \succ$ just
expresses the invariance of the correlator under global SU(2)
transformations.
Since the boundary
 operator
$\psi$
 is quasiprimary\cite{BPZ},
it
is annihilated
by $L_1$. Acting with $L_1$
 on
(\ref{martingale},\ref{diffusion-F},\ref{Hdiffusion})
yields
the first order
differential equation
\begin{align}
\label{KZlevel1} \left( L_{-1} - 2 \gamma {J}^a_{-1} J^a_0 \right
) \! \prec \! {\cal F} \! \succ &= 0,
\end{align}
where $2\gamma\! =\! \tau/[6\! -\! (2h_0\!+\!1)\kappa]$.\!
Acting\! with\! $L_1$ again  yields\!
\begin{align}
\left (L_0 - \gamma J^a_0 J^a_0 \right ) \! \prec \! {\cal F} \!
\succ &= 0, \label{KZlevelzero}
\end{align}
which gives $h_0 = \gamma s(s+1)$, with $s = 1/2$ the spin at the
tip.
We recognize
in (\ref{KZlevel1},\ref{KZlevelzero})
the first two modes of the Sugawara
relation\cite{Knizhnik:Zamolodchikov}.
Eq. (\ref{KZlevel1}) is the Knizhnik-Zamolodchikov equation
arising as a level 1 Null vector of the operator at the tip, and
(\ref{KZlevelzero})
yields
the familiar conformal weight of the WZW
model \cite{Knizhnik:Zamolodchikov}.

Finally, if we parametrize $\gamma ={1/(k+2)}$, and use the
relationship (\ref{kappa+tau}) between $\kappa$ and $\tau$, we
obtain for $k \neq 1$
\begin{align}
\tau &= \frac{4}{k+3}, & \kappa = 4\frac{k+2}{k+3}.
\label{TauKappa-k}
\end{align}
The above conditions, however, do not specify $\tau$ and $\kappa$,
at $k=1$. The reason for this is that at $k = 1$ the WZW model is
a CFT with $c = 1$, and can be equivalently described in terms of
Abelian fields (see below).

In the following paragraph, the parameter $k$ will be identified
with the level of the $\text{su}(2)_k$ current algebra.

\paragraph*{ Null vectors.}

The action of the operators (\ref{KM-modes}) creates descendant
operators of the
boundary operator
$\psi$
at the tip.
Eqs.
(\ref{martingale},\ref{diffusion-F},\ref{Hdiffusion})
mean
 that $\chi = H \psi$ is
a Null vector \cite{Knizhnik:Zamolodchikov, BPZ}. Acting with
${\cal J}_1^b$
on (\ref{KZlevel1})
shows that the parameter $k$
defined in the last section is the level of the Kac-Moody algebra.
Furthermore, acting with ${\cal L}_2$
on
(\ref{martingale},\ref{diffusion-F},\ref{Hdiffusion})
expresses
 the central charge as $c={3 \over 4} \tau k +(3 \kappa
-8) h_0$. Inserting $h_0 = (3/4)/(k+2)$ into the equation
appearing in the text below (\ref{KZlevel1}), yields a quadratic
equation for the level $k$ as a function of $\kappa$ and $\tau$.
Thus, both $c$ and $k$ are functions of $\kappa$ and $\tau$. Using
(\ref{kappa+tau}) one
 recovers $c=3k/(k+2)$,
the familiar
central charge of $\text{su}(2)_k$.

\paragraph {Relation to unitary minimal models.}

As in
usual SLE,
the variance $\kappa$ alone determines the
{\it geometry}
 of the trace. The CFT that describes the geometry of the trace
alone corresponds to SLE$_\kappa$, with $\kappa$ given by
(\ref{TauKappa-k}). The corresponding portion of central charge is
$c_\kappa = 1 - \frac{6}{(k+2)(k+3)}$.
We observe that this is the central charge of the minimal unitary
model, which can be thought of as the coset model $\text{su}(2)_k
\oplus \text{su}(2)_1/\text{su}(2)_{k+1}$. The remaining part of
the central charge, $c_\tau = 3(k + 1)/(k + 3) - 1$, is the same
as that of the coset $\text{su}(2)_{k+1}/\text{u}(1)_{k+1}$ or
$Z_{k+1}$ parafermion\cite{parafermion}. It describes
the `{\it twisting}' of
the trace.

\paragraph*{Abelian reduction.}

Our stochastic approach is easily specialized to the case of the
abelian group U(1). In this case the requirement of Fuchsian
singularities that led to Eq. (\ref{kappa+tau}) gives the simple
condition $\kappa = 4$ and eventually leads to $c = 1$.
Eqs.  (\ref{martingale},\ref{diffusion-F},\ref{Hdiffusion})
still hold
but have a
different meaning. They are satisfied by a correlator at $c = 1$,
where the boundary operator  has dimension $4h = 1 - \tau$. By
varying $h$ away from $h=1/4$ ($\tau=0$), corresponding to the
SU(2)  WZW model at level $k=1$, one obtains an evolution which
depends both on the geometry and on the U(1) current algebra
symmetry. ---
An analog
 of
(\ref{martingale},\ref{diffusion-F},\ref{Hdiffusion})
for the
Abelian
case was obtained in Ref. \cite{228:Cardy} using methods very
different from ours.

\paragraph*{Acknowledgement:}

We benefitted from discussions with M. Bauer, D. Bernard, J.
Cardy, B. Duplantier, L. Kadanoff, I. Kostov, B. Nienhuis, and I.
Rushkin. This work was supported in part by the NSF under
DMR-00-75064 (A.W.W.L.), DMR-0220198 (PW). EB, IG and PW were
supported by NSF MRSEC Program under DMR-0213745. I. A. G. was
supported by the Alfred P. Sloan Foundation, the Research
Corporation and NSF Career award DMR-0448820.


\begin{thebibliography}{10}
\expandafter\ifx\csname
natexlab\endcsname\relax\def\natexlab#1{#1}\fi
\expandafter\ifx\csname bibnamefont\endcsname\relax
   \def\bibnamefont#1{#1}\fi
\expandafter\ifx\csname bibfnamefont\endcsname\relax
   \def\bibfnamefont#1{#1}\fi
\expandafter\ifx\csname citenamefont\endcsname\relax
   \def\citenamefont#1{#1}\fi
\expandafter\ifx\csname url\endcsname\relax
   \def\url#1{\texttt{#1}}\fi
\expandafter\ifx\csname
urlprefix\endcsname\relax\def\urlprefix{URL }\fi
\providecommand{\bibinfo}[2]{#2}
\providecommand{\eprint}[2][]{\url{#2}}








\bibitem{BPZ} A. A. Belavin, A. M. Polyakov, and A. B.
Zamolodchikov, Nucl. Phys. {\bf B241}, 333 (1984).

\bibitem{nienhuis} B. Nienhuis, in {\it Phase transitions
and critical phenomena}, Vol. 11, Academic Press, London, 1987.

\bibitem[{\citenamefont{Schramm}(2000)}]{Schramm:SLE}
\bibinfo{author}{\bibfnamefont{O.}~\bibnamefont{Schramm}},
   \bibinfo{journal}{Israel J. Math} \textbf{\bibinfo{volume}{118}},
   \bibinfo{pages}{221} (\bibinfo{year}{2000}).

\bibitem{Lawler book} G. F. Lawler, {\it Conformally invariant
processes in the plane}, American Mathematical Society, 2005; W.
Werner, "Random planar curves and Schramm-Loewner evolutions", in
{\it Lecture Notes in Mathematics}, {\bf 1840}, Springer, 2004.

\bibitem{Cardy Review}For a recent review see:
  J.~Cardy,
  Annals Phys.\  {\bf 318}, 81 (2005)
  cond-mat/0503313.




\bibitem[{\citenamefont{Bauer and Bernard}(2003)}]{Bauer:2002tf}
\bibinfo{author}{\bibfnamefont{M.}~\bibnamefont{Bauer}} \bibnamefont{and}
   \bibinfo{author}{\bibfnamefont{D.}~\bibnamefont{Bernard}},
   \bibinfo{journal}{Commun. Math. Phys.} \textbf{\bibinfo{volume}{239}},
   \bibinfo{pages}{493} (\bibinfo{year}{2003}); \eprint{cond-mat/0412372}.


\bibitem[{\citenamefont{Knizhnik and
   Zamolodchikov}(1984)}]{Knizhnik:Zamolodchikov}
\bibinfo{author}{\bibfnamefont{V.~G.} \bibnamefont{Knizhnik}} \bibnamefont{and}
   \bibinfo{author}{\bibfnamefont{A.~B.} \bibnamefont{Zamolodchikov}},
   \bibinfo{journal}{Nucl. Phys.} \textbf{\bibinfo{volume}{B247}},
   \bibinfo{pages}{83} (\bibinfo{year}{1984}).


\bibitem{SpinChains}
\!I.\! Affleck\!  and\! F.\! D.\! M.\! Haldane,\! Phys.\! Rev.\!
{\bf B36},\! 5291\! (1987).\!


\bibitem{Nozieres+Blandin}P. Nozi\`eres and A. Blandin, J. Phys.
Paris {\bf 41}, 193 (1980).

\bibitem{Alekseev:1998mc} \!A.\! Y.\! Alekseev\! and\! V.\! Schomerus\! Phys.\! Rev.\!
D60,\! 061901\! (1999);\! G.\! Felder et al.,\! J.\! Geom.\!
Phys.\! {\bf 34},\! 162\! (2000); \! V. Schomerus,\! Class.\!
Quantum\! Grav.\! {\bf 19},\! 5781\! (2002).\!


\bibitem{kondev} J. Kondev, C. L. Henley, Nucl. Phys. {\textbf{B464}},
540 (1996), J. Kondev, Int. J. Mod. Phys. {\textbf{B11}}, 153
(1997); N. Read, reported at Kagom\'e Workshop (Jan. 1992),
unpublished.

\bibitem{Rasmussen} A very different approach was suggested in
J. Rasmussen, hep-th/0409026.

\bibitem[{\citenamefont{Cardy}(1984)}]{Cardy1984}
\bibinfo{author}{\bibfnamefont{J.~L.} \bibnamefont{Cardy}},
   \bibinfo{journal}{Nucl. Phys.} \textbf{\bibinfo{volume}{B240}},
   \bibinfo{pages}{514} (\bibinfo{year}{1984}).

\bibitem[{\citenamefont{Ishibashi}(1989)}]{Ishibashi:1988kg}
\bibinfo{author}{\bibfnamefont{N.}~\bibnamefont{Ishibashi}},
   \bibinfo{journal}{Mod. Phys. Lett.} \textbf{\bibinfo{volume}{A4}},
   \bibinfo{pages}{251} (\bibinfo{year}{1989}).

\bibitem[{\citenamefont{Affleck and Ludwig}(1991)}]{AffleckLudwig1991NuclPhysB}
\bibinfo{author}{\bibfnamefont{I.}~\bibnamefont{Affleck}} \bibnamefont{and}
   \bibinfo{author}{\bibfnamefont{A.~W.~W.} \bibnamefont{Ludwig}},
   \bibinfo{journal}{Nucl. Phys.} \textbf{\bibinfo{volume}{B352}},
   \bibinfo{pages}{849} (\bibinfo{year}{1991});
   \bibinfo{journal}{Nucl. Phys.} \textbf{\bibinfo{volume}{B428}},
   \bibinfo{pages}{545} (\bibinfo{year}{1994}).

\bibitem{CardyNPB1989} J. L. Cardy, Nucl. Phys. {\bf B324}, 581
(1989); I. Affleck and A.~W.~W. Ludwig, J. Phys. {\bf A27}, 5375
(1994).


\bibitem{parafermion} \! A.\! B.\! Zamolodchikov,\! V.\!A.\! Fateev,\! JETP {\bf 62}\! (1985),\! 215.\!


\bibitem[{\citenamefont{Cardy}(2004)}]{228:Cardy}
\bibinfo{author}{\bibfnamefont{J.}~\bibnamefont{Cardy}}
(\bibinfo{year}{2004}),
   \eprint{math-ph/0412033}.


\end{thebibliography}
\end{document}